# Comet Machholz (C/2004 Q2): morphological structures in the inner coma and rotation parameters.

**Federico Manzini\*, Raoul Behrend###, Lorenzo Comolli\*\*, Virginio Oldani\*, Cristiano B. Cosmovici#, Roberto Crippa\*\*\*, Cesare Guaita\*\*, Gottfried Schwarz\*\*\*\*, Josep Coloma##**

\* SAS, Stazione Astronomica, Sozzago, Italy (IAU A12)

\*\* GAT, Tradate, Italy

\*\*\* Osservatorio di Tradate, Italy (IAU B13)

\*\*\*\* DLR, Wessling, Germany

# IFSI / INAF, Roma, Italy

## Observatorio de Sant Gervasi (IAU A90)

### Geneva Observatory, Switzerland

**Dr. Federico Manzini**

Stazione Astronomica

28060 SOZZAGO (NO) – ITALY

phone: +39 340 8077664

fax : +39 02 97289464

e-mail: manzini.ff@aruba.it

**Abstract:**

Extensive observations of comet C/2004 Q2 (Machholz) were carried out between August 2004 and May 2005. The images obtained were used to investigate the comet's inner coma features at resolutions between 350 and 1500 km/pixel.

A photometric analysis of the dust outflowing from the comet's nucleus and the study of the motion of the morphological structures in the inner coma indicated that the rotation period of the nucleus was most likely around 0.74 days.

A thorough investigation of the inner coma morphology allowed us to observe two main active sources on the comet's nucleus, at a latitude of $+85° \pm 5°$ and $+45° \pm 5°$, respectively. Further sources have been observed, but their activity ran out quite rapidly over time; the most relevant was at $lat_{com.} = 25° \pm 5°$.

Graphic simulations of the geometrical conditions of observation of the inner coma were compared with the images and used to determine a pole orientation at RA=$95° \pm 5°$, Dec=$+35° \pm 5°$.

The comet's spin axis was lying nearly on the plane of the sky during the first decade of December 2004.

**Keywords:** Comets, comet Machholz, Jets, Rotation Period, Shells.

## 1. INTRODUCTION.

Comet C/2004 Q2 (Machholz) was discovered by Don Machholz (*Machholz, 2004*) in the morning of August 27, 2004. It seems to be a member of the Oort cloud, with an estimated orbital period of 113,000 years (*Marsden, 2005*).

Due to the favorable geometrical conditions of the appearance (*Green, 2004*), the comet became as bright as magnitude 3.5, reaching its highest luminosity in the first week of January 2005, when it passed at the perigee (*Yoshida, 2006, and Yoshida, priv. comm.*). Moreover, comet Machholz became circumpolar for the northern observers in March and April 2005, thus allowing extensive observational sessions.

On February 9, 2005, the first determinations of the rotation period and of the pole position were announced by Sastri (*Sastri et al., 2005a, b*), but his solution (rotation period of 9.1 h, pole orientation at RA=+190°, Dec=+50°) was demonstrated to be incorrect by Farnham and co-workers (*Farnham et al., 2007b*), who proposed a new estimate for the rotation period of 0.733 days (17.6 h) and for the pole position at RA=+50°, Dec=+35°, after observing the comet extensively from February to April 2005.

The determination of the rotation period and the geometrical properties of the nucleus of a comet is an extremely complex task. They can be estimated on the basis of morphological studies, but only through a continuous monitoring of the comet's jets or shells (if observable), and are dependent on their geometry. In some cases, differential photometry may be applied to images taken during prolonged observations obtained over several consecutive nights.

The study of the morphology of the inner coma, as well as of the variations over time of the observed structures, provides important information to understand the evolution of the active areas on the nucleus. In the recent past, the study of the evolution of jets and shells in the inner coma of comet Hale-Bopp allowed us to discover a close correlation with the rotation of the nucleus (*Manzini et al., 1997, Schwarz et al., 1997*). Comet Ikeya-Zhang was also found to be particularly suitable for similar morphological studies (*Manzini et al., 2006*).

Dust features are well visible inside the inner coma of the most active comets, since they show a higher signal-to-noise ratio than gas features. Moreover, dust structures can be observed up to a long distance from the nucleus, as dust is less affected than gases by the solar radiation. However, molecular dissociation of gases occurs very close to the nucleus, and this process causes important variations in the wandering speed which may mask the position of the active sources.

Observations made with broadband filters or without filters are dominated by the reflection of sun light on dust in the inner coma: this effect increases the signal-to-noise ratio for dust, thus minimizing the contribution due to the presence of gases (*Farnham et al., 2007*).

The objectives of our work can be summarized as follows:

1. observation and characterization of the details of the inner coma through image enhancement by means of different filters;
2. differential photometry of the brightness of the false nucleus and of the inner coma
3. determination of the rotation period and of the orientation of the spin axis of the nucleus.

## 2. METHODS OF OBSERVATION.

Different series of images of comet Machholz were taken by our group at several observatories mainly located in Northern Italy (**Table 1**).

The instruments used permitted a sufficient spatial resolution to investigate the morphology of the inner coma, and had been successfully used for previous studies on comets Hyakutake, Hale-Bopp (*Manzini et al., 1997, Schwarz et al., 1997*) and Ikeya-Zhang (*Manzini et al., 2006*).

The first long series of images of comet Machholz were taken on December 10, 2004; the geometrical conditions became favorable only around that date, when the comet left the constellation Columba (very low at our latitudes) to rise in declination.

The comet was extensively observed over several consecutive nights from 10-13 December 2004, 1-6 and 10-16 January 2005. Ten additional observing sessions were spread between August 2004 and May 2005. All observations were carried out under very good atmospheric conditions.

The favourable geometrical conditions of the appearance allowed us to obtain instrumental resolutions of up to about 300 km/pixel during the first decade of January, 2005.

The full list of our images, by date and with full description of their characteristics, is shown in **Table 2**.

### 3. IMAGE PROCESSING.

All CCD images were first processed by applying conventional methods of data reduction (bias, dark and flat-field images, collected on the same night of the observing sessions).

When studying the coma morphology, image enhancements are necessary to reduce the bright central peak at the optocenter and to maximize the signal-to-noise ratio. Whenever possible, we treated co-added images taken on the same night, instead of processing every single one. The time lag between our co-added images never exceeded 30 min. These frames were registered to a common center (the brightest peak of the optocenter) at a sub-pixel level.

We validated this procedure by verifying that all the details in the co-added images were the same as, and in the same position of, those in every single frame; therefore, no smearing effect was evident.

Further processing was performed by means of the image processing techniques described below.

#### 3.1 Larson-Sekanina algorithm.

All the images collected were processed by means of the Larson-Sekanina spatial filter (*Larson and Sekanina, 1984*) to highlight radial features (i.e. jets) and haloes (i.e. shells).

#### 3.2 Unsharp masking.

To show at the same time all the features in the inner coma of comet Machholz, such as clumps of outflowing material, haloes and jets, we used a standardized process based on the following steps:

- typical unsharp masking, after treating the dividing image with a low-pass filter with a 5-sigma radius,
- removal of the median radial gradient centered on the brightest peak of the optocenter in the comet's nucleus.

To avoid false interpretations of the different series of images, we always used the same processing parameters for all images.

#### 3.3 Ring masking technique.

The images were processed with two algorithms:

- a ring masking technique to highlight radial features (*i.e.* jets)
- a radial de-trending technique to enhance shells, haloes and bow shock-like structures.

Our ring masking technique is based on the idea that any radial outflow (i.e. a jet) leads to slightly enhanced levels of local radiance, which however only account for a small percentage of the mean intensity, and are therefore totally invisible without additional image processing.

In order to detect them, we first compare the radiance levels along concentric rings around the nucleus: in the case of a perfect symmetry, no differences in intensity are noticed; on the contrary, if local jet structures are present, they determine small localized differences in the intensity levels, which may then better be visualized with appropriate image processing techniques.

To increase the visibility of these small differences, we apply the ring masking technique transforming the images into polar coordinates around the center of brightness (i.e. the assumed position of the nucleus).

The ring masking technique was successfully applied previously to comets Halley and Hale-Bopp (*Cosmovici et al. 1988, 1993; Schwarz et al. 1989, 1997*).

The results of the above processing steps are shown in ***Figure 1*** for both January 2 and 6, 2005:

- Radial structures departing from the nucleus are very evident and the outflow of material can be easily observed despite the time span between two images in the same day is only about 3 hours.

- No halo-like structures are instead visible in any of our images, whereas, if present, the radial de-trending technique would have evidenced them clearly.

### 3.4    Graphic simulations.

Graphic simulations of the geometrical conditions of observation of the nucleus were compared with images taken within a long time between them.

Furthermore, simulations of the position of the rotation axis of the nucleus were compared with the results obtained through image enhancement to verify their correspondence to the model.

All graphic simulations were obtained with the software Starry Night PP v. 5.8.4 (Imaginova Canada Ltd., Toronto, CDN).

## 4. COMA MORPHOLOGY AND ANALYSYS OF THE OBSERVED FEATURES.

Asymmetries in the coma were already visible in our reduced CCD images, even without any enhancement; their presence indicated a strong activity on the comet's nucleus, probably limited to few areas.

Coma features have been observed in other comets, and can be used to determine the properties of the nucleus. (*Chen et al., 1987; Yoshida et al., 1993; Samarasinha et al., 1997; Samarasinha, 2000*). Details inside the inner coma were systematically observed also in the case of comet C/2004 Q2 (Machholz) (*Farnham et al., 2007; Lin et al., 2007*).

Since these features may look different at different resolutions, we decided to analyze only the images obtained at the SAS Observatory in order to create a sort of 'atlas' of all the observed features. These images were successively compared with those taken at the other observing sites.

### 4.1    Dec. 10, 2004 (R = 1.39 AU, delta = 0.51 AU; image resolution: 500 km/pixel).

The inner coma morphology was characterized by four main jets coming up near the Position Angles (PA) 10°, 140°, 210° and 270° (all data are given with a precision of ±5°), and by some small fainter jets (*Figure 2a*, right). Since the comet was almost in opposition to the Sun, its tail was developing towards North: thus, the jet in PA 10° is likely attributable to the presence of the tail.

Several previous observations and published studies have shown that the presence of rectilinear jets of similar brightness inside the coma of a comet, if seen from a low phase angle, is the consequence of the outflow of material from active areas located on the comet's nucleus; the number of these active sites is half that of the jets that appear in symmetrical position to the rotation axis.

Two similar, rectilinear and symmetrical 'porcupine' jets were clearly visible in PA 140° and 270°: their axis of symmetry was likely to coincide with the projected direction of the rotation axis of the comet. They were probably produced by a single active area located at low latitudes on the nucleus: $\lambda = 90° - 0.5 * (270° - 140°) = 25° ± 5°$

A further rectilinear 'porcupine' jet in PA 210°, without a symmetrical counterpart, could be due to a single active area located very close to one of the rotation poles of the comet. In fact, if the jet had not been at a short distance from the rotation pole, a progressively larger fan structure would have been observed in relation with its distance from the pole (*Sekanina, 1998*).

The observation of these features supported the initial assumption that the rotation axis was lying nearly on the plane of the sky, with a small inclination (probably between +20° and -20°), and that its direction passed between PA 30° and PA 210°, through the southern jet.

**Figure 2a** shows the features in the inner coma of comet Machholz; the black and white streaks are star trails. A spherical simulation of the rotation state of the comet's nucleus, consistent with the above hypotheses, is shown on the left side (not in scale with the picture on the right side); the arrows indicate the positions of the jets visible in the image, put on a grid of comet coordinates. The simulation is here represented with a direction of the rotation axis towards RA=95° and Dec=+35°.

The tail outflow is not shown. The simulation shows also the effect of the sun light on the nucleus, and the projected sunward direction is marked.

The polar jet would have been in a nearly sub-solar position during the 'comet day', thus producing a continuous outflow, whereas the second jet, positioned at the derived latitude of 25°, might have reached the 'comet night' during the rotation of the nucleus. However, according to this simulation, this would have happened in an area which was hidden to our view.

### 4.2 Dec. 20, 2004 (R = 1.33 AU, delta = 0.42 AU; image resolution: 416 km/pixel).

Two jets, directed respectively towards PA 270° and 250°, looked very bright; the first was already known, the second showed a very faint counterpart with respect to the rotation axis determined previously. This was already visible on December 10, but only after applying an intense stretching; it was increasingly evident in the images of the following days (December 11 to 14, and 18) (**Figure 2b**).

It is likely that on those days a new site became active on the comet's nucleus: several features would have originated from this source during the following days.

Unfortunately, due to bad weather conditions over our sites, we could not follow the evolution of this jet from December 21 to December 31.

### 4.3 Jan. 01, 2005 (R = 1.26 AU, delta = 0.35 AU; image resolution: 345 km/pixel).

The comet had rapidly come closer to the Earth and remained at nearly the same distance for about 10 days. Three major jets were still visible in PA 250°, PA 200° and PA 290°; the first held a position corresponding to that of the jet defined as "polar" on December 10. The others were in symmetrical position to the spin axis determined previously; they probably originated from that new active source, located at a latitude of 45°±5°, which had already been observed on December 20. The active area that had produced a jet at latitude 25°±5° had ceased its activity (**Figure 2c**).

The graphic simulation of the nucleus for that date, obtained once again with a spin axis directed at RA=95° and Dec=+35°, shows the same hemisphere observed on December 10 still brightened by the sun light; the angle between the rotation axis and the plane of the sky is wider than 40°.

The polar jet is very active and, according to our simulation, it is in sub-solar position, constantly hit by the solar radiation; the jet located at mid-latitude is well visible on both sides of the axis of symmetry: the simulation shows that their originating site is never subjected to "setting".

***4.4     Jan. 13, 2005 (R = 1.22 AU, delta = 0.36 AU; image resolution: 360 km/pixel).***

According to our simulation (**Figure 2d**), the rotation axis was directed towards PA 260°; the angular variation in the PA of the jets can be compared with that of the previous date on the reference image.

The emitting activity modulated by the rotation of the active source located at a latitude of 45°±5° can be observed in images processed to increase as much as possible the faintest features on the jets. The polar jet still appears here in sub-solar position. It showed a variable intensity in our animations (*http://www.foam13.it/new/machholz/mchoolz.htm*), as well as in other images taken on January 11, 14 and 15, 2005. The apparent clump of material visible on the polar jet is however not real, as it is the result of an increase in brightness consequent to the perspective sum of material outflowing from the polar jet with that outflowing from the rotating jet located at mid-latitudes.

Shell structures are not visible at this resolution, because the angle between the spin axis and the plane of the sky is quite small and the field of view of the image is wide.

**Figure 3** has been stretched to show the emission of material modulated by the rotation; the positions of "clumps" of material drifting away from the nucleus, changing shape and fading due to a gradual loss of dust, are indicated; the two images were taken 130 minutes apart.

***4.5     Feb. 08, 2005 (R = 1.23 AU, delta = 0.56 AU; image resolution: 480 km/pixel).***

The comet had drifted far away from the Earth and the Sun, and the geometrical conditions of the observation had changed with respect to the previous sessions.

The simulation in **Figure 2e** shows that the rotation axis was now directed towards PA 290°, with a greater solar radiation on the hemisphere hidden to our view. In such conditions, the polar jet should have been still very active, whereas the jet located at mid-latitude was continuously experiencing 'day' and 'night'.

The images are consistent with the simulation; the jet at intermediate latitude was overwhelming where it was affected by the radiation of the Sun, whereas the emission in symmetrical position was very scarce: in that case the jet was in umbra and was therefore almost inactive.

Along the axes of the jets the presence of aggregates of material modulated by the rotation was still observable, appearing like 'plumes'.

***4.6     Mar. 19, 2005 (R = 1.46 AU, delta = 0.96 AU; image resolution: 950 km/pixel).***

The apparent direction of the comet's spin axis had turned by 180° in four months. The geometrical conditions of the observation had changed and the rotation axis, according to the simulation, was only slightly inclined on the plane of the sky (**Figure 2f**).

In the images, the jets had become straight and no effects due to the rotation of the nucleus (such as clumps of material) were visible.

The polar jet was still very active; in the simulation it was enlightened for the entire comet's day,

although it was no longer in sub-solar position.

The jet at mid-latitude was more intense where it was illuminated by the Sun; according to the simulation, it had moved to a sub-solar position, but in the symmetrical position it was virtually inactive because it was in the night side of the comet.

The images processed by means of the *ring masking* technique show an intense outflow of material resulting from the jet at mid-latitude oriented towards PA 350°, and a small curved jet towards PA 80°; it is likely that this structure was the symmetrical counterpart of the main jet. It is hardly visible with other image processing techniques.

### 4.7    Apr. 18, 2005 (R = 1.74 AU, delta = 1.28 AU, image resolution: 2330 km/pixel).

The simulation shows that the comet's spin axis was directed towards PA 90°, the polar regions received little solar radiation, whilst the Sun was still present at mid-latitudes. Since the comet was going far away, the images showed a quite low resolution so that details could be observed only at a medium distance from the nucleus.

The polar jet was nearly extinguished: it presented a low signal and could be seen highlighted against a noisy background only after using an extreme stretching.

The jet at mid-latitude was still very active where it was hit by the solar radiation. The jet in symmetrical position was invisible because it was turned off: in fact, it was in the 'night side' of the nucleus (**Figure 4**).

### 4.8    Animations.

The images obtained from January 1 to 6, 2005 were processed by means of the *radial de-trending* technique, and then rotated to keep the direction of the comet's rotation axis fixed. They were then phased to a period of 17.7 h, and assembled to obtain an animation of the outflow of material from the nucleus (the animation can be found at http://gruppoastronomicotradatese.it/comete/machholz.htm). Since the images were not taken on the same day, the outflow of material is only demonstrative of the dust motion in the inner coma.

## 5.  VERIFICATION OF THE SPIN AXIS DIRECTION.

A very interesting paper on comet Machholz has been published by Farnham and co-workers (*Farnham et al., 2007*).

The authors carried out extensive observations of the comet's inner coma in March and April 2005 at high resolutions (between 125 and 165 km/pixel) with the 2.1 m Kitt Peak National Observatory (KPNO) telescope (AZ, USA). Images were obtained using narrowband filters to isolate the CN features. The published images, as well as the animations in the website indicated in the same paper (http://www.astro.umd.edu/farnham/Machholz), show the presence of CN gas outflowing from the nucleus with a characteristic corkscrew pattern. According to the authors, the nucleus had a rotation period of 17.6 ± 0.05 h and a pole orientation at RA=50° and Dec=+35°.

In order to complete our verification phase, we introduced the values of the spin axis direction proposed by Farnham in the simulation software and compared the results with our observations; the highest-resolution images, taken on January 1 and 13, 2005, were selected for this verification. According to the simulations, the comet's rotation pole should have been pointing towards the direction of observation on those days (***Figure 5,*** right). Therefore, even considering the presence of only a single active area, full shells or segments of Archimedean spirals should have been observed around the nucleus.

None of these features is present in our images treated with the methods described above; on the contrary, only jets with nearly rectilinear structure are visible. In particular, we took great care in applying the radial de-trending technique to enhance shells, but we found no evidence of them (***Figure 5,*** left).

Therefore, we believe that the direction for the rotation axis proposed by Farnham is not correct.

Conversely, the results obtained performing a simulation of the geometric conditions of the observation of the nucleus with the values that we found (RA=95°±5° and Dec=+35°±5°) are consistent with what we observed on our images.

## 6. ROTATION PERIOD.

### 6.1 Analysis through observation of the displacement of dust features.

The determination of the rotation period of comet Machholz was obtained through the analysis of the images taken on January 14, 2005 at the GAT Observatory (GAT#1, **Table 1**) over more than 6 hours. On these images we could observe two haloes of which we measured the expansion from 18h32m to 23h59m UT. The two halos were actually part of a corkscrew-shaped feature that was outflowing from the comet's nucleus, as derived comparing the images with the simulations.

The analysis of the motion of similar structures allows to determine the rotation period of a comet; we had previously applied this method to estimate the rotation period of comets Hale-Bopp and Ikeya-Zhang (*Manzini et al., 2001, 2006*).

To perform the study of the images, we made a photometric analysis of two lines drawn across the comet's nucleus and the two brightest features of the haloes in two images taken at 18h52m UT and 23h19m UT (**Figure 6**). It was thus possible to measure for each of the two outflowing features the mean separation **L** in pixels (projected on the plane of the sky) between the two images and then, the distance of the comet being known, to convert it into kilometers.

Assuming that a new feature had formed after every rotation of the nucleus, the rotation period **P**, expressed in days, is simply related to the expansion velocity **V** (km·day$^{-1}$) of the features (*Braunstein et al., 1997*) by the relationship: $\boldsymbol{P = L \cdot V^{1}}$.

The "projected" expansion velocity was not affected by the orientation of the comet's rotation axis towards the Earth.

The mutual distance, projected on the plane of the sky, between the first two observable haloes on our images was 13,900 ± 400 km at the distance of the comet. The projection on the plane of the sky of the expansion velocity, as measured on the images taken on January 14, was 785 ± 45 km/h (0.218 ± 0.013 km/s). Through these data we could estimate a comet's rotation period of 0.74 ± 0.03 days (17.8 ± 0.7 h).

Further confirmations of this value for the rotation period were also obtained in the same way from images taken earlier, on January 3 and 4, 2005.

### 6.2 Analysis through photometry.

Comet Machholz was extensively observed at SAS Observatory (**Table 1**) over several hours on the nights of January 3-4, 4-5, 6-7, 11-12 and 13-14 with the aim to determining any photometric variations of the inner coma. A further observation session was performed on the night of January 10 in the OSG Observatory; the geometrical conditions of these observation sessions are shown in Table 3.

All frames were analyzed by means of differential photometry with respect to non-variable stars included in the field of view; we used the same stars on the same night whenever possible (although this was not always feasible due to the rapid motion of the comet within the field of view). The use of differential photometry with four reference stars always present in every frame allowed us to obtain measures with a precision of ±0.03 magnitudes in every observing session, with the exception of the night of January 13, 2005, during we could obtain a precision of ±0.04 magnitudes.

The instrumental magnitudes on the images taken at SAS Observatory were measured within a 10-pixel radius circular window (corresponding to 13.6 arcseconds, or about 3500 km at the comet distance) centered on the comet's nucleus. At the same time, measurements were made within a concentric annulus of 30-pixel (40.6 arcseconds) and 40-pixel (54.4 arcseconds) inner and outer radii, respectively, to detect the sky level; the median value of all pixels comprised within the annulus was calculated to eliminate any effect due to the coma and the tail. A further test made with a 90-pixel and a 100-pixel inner and outer radii (where the effect of the coma and the tail was minimal) did not show significant variations compared with the values that we identified to be optimal.

The size of the central circular window allowed us to take into account any variations in the brightness due to the features placed on the nucleus and, at the same time, to new material emitted from the active sites on the nucleus. In fact, jets developing from active areas on the comets' nuclei become clearly visible after they have passed the point of sunrise and have been exposed to the sun; the jets produce a lot of material during the nucleus rotation until they reach the point of sunset on their horizon, when they cease any activity.

A photometric analysis of a small area around the nucleus prolonged for several hours may therefore show the jet emission through an increase in the brightness of the area analyzed. On the contrary, a decrease in brightness is observable when the jets are in the dark side of the nucleus.

This method seems to be efficient for the analysis of active comets that present few emitting jets and in particular of those with a near-Earth orbital pattern, as it was the case for comet Machholz.

All photometric measurements from January 3 to 14, 2005, were merged at the Geneva Observatory in the light curve shown in **Figure 7**. The best estimate of the rotation period that could be derived from this analysis was P=0.742 ± 0.005 days (17.49 ± 0.12 h). In **Figure 8** the same data have been phased with the estimated period.

## 7. DISCUSSION.

The analysis of the morphology of the inner coma of comet C/2004 Q2 (Machholz) allowed us to determine some properties of its nucleus, such as the approximate location of active areas, the direction of the spin axis and the rotation period.

The jets observed in the dust coma during the period of our observations could be explained by the presence of three active regions on the surface of the nucleus. Simulations of the geometric conditions of the observation provided an explanation for the features observed in the inner coma, considering the presence of one active site located near the polar region ($lat_{com.} = 85° ± 5°$) and the others at mid or low latitudes ($lat_{com.} = 45° ± 5°$ and $lat_{com.} = 25° ± 5°$).

The positions of the active areas that were emitting dust and producing the jets observed in our images seem to be different from the sources that were releasing gases (*Farnham et al., 2007b; Lin et al., 2007*).

Fitting our images with our model allowed us to derive a direction of the spin axis of the nucleus at RA=95°±5° and Dec=+35°±5°. This estimation is significantly different from the value of RA=50° obtained by Farnham (*Farnham et al., 2007b*); in fact, performing a simulation of the geometric conditions of the observation of the nucleus with this value, the results obtained are not in agreement with what is observed on the images.

Using of two different methods, we obtained comparable values for the rotation period:

- 0.740 ± 0.03 days, through a "dynamical" method, by observing the wandering of dust features from the nucleus;
- 0.742 ± 0.005 days, through a photometric method, after prolonged observing sessions.

These results are also in agreement with those proposed by Farnham.

In a recent paper published by Reyniers (*Reyniers et al. 2009*), a period of 9.1 h was found by means of CCD photometry. However, comparing their results with images obtained in the narrowband CN filter, the authors did not exclude the possibility that their method sampled P/2 instead of P. In the latter case, a period of 18.2 h (0.758 days) would be in perfect agreement also with our estimate and would provide a further confirmation of the validity of the method applied.

We may therefore conclude that:

- the rotation of the nucleus of comet Machholz occurs on a single axis and is not chaotic;
- the rotation period is most probably within the above indicated range, i.e. between 0.74 and 0.75 days.

Unfortunately, our measurements do not allow estimating possible precession effects.

**Acknowledgements**

We would like to thank:

Dr. H. Mikuz, Crni Vrh Observatory (IAU 106), for providing the useful picture of April 18, 2005, and dr. S. Yoshida for providing the magnitude vs. time graph of the complete apparition of comet Machholz.

A. Brunati for the great help given in taking images.

A. Dimai (Cortina Astronomical Observatory), P. Recaldini (Aosta Valley Astronomical Observatory), and W. Borghini (Osservatorio Astronomico di Casasco) for providing images.

Jana Pittichova for the suggestions provided, which allowed us to improve this paper significantly.

**REFERENCES.**

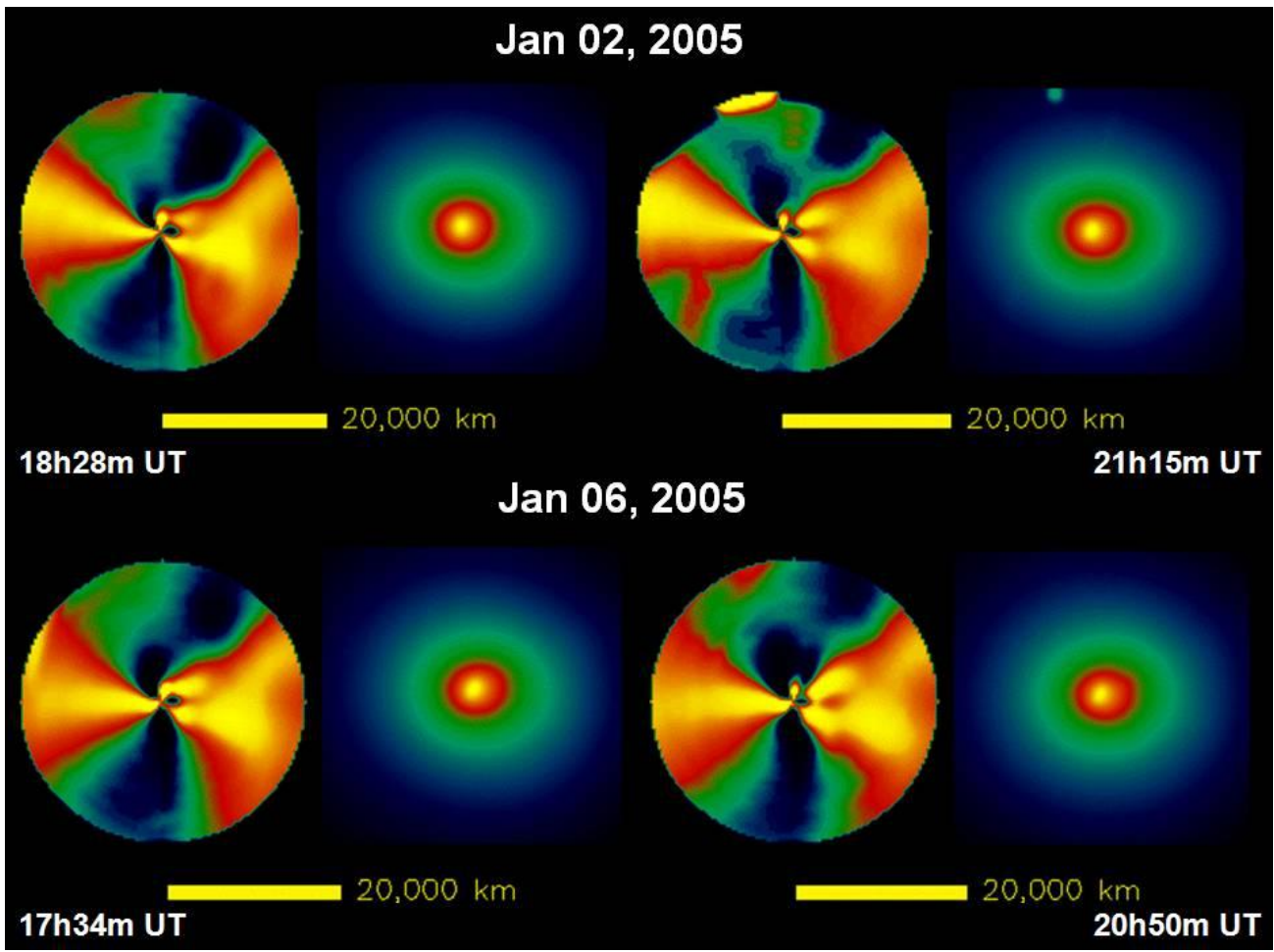

**Fig. 1**

The results of image processing by the *ring masking* and the *radial de-trending* technique are shown in the first and third column. Images were taken on January 2 (mean timed at 18h28m and 21h15m UT) and January 6 (mean timed at 17h34m and 20h50m UT), 2005. The outflow of material from active sources is visible in form of knots, superimposed on the jets right to the nucleus, while the tail of comet Machholz is on the left. Spot on the top and dotted trails are due to field stars.

In the second and fourth column unprocessed, logarithmically scaled color-coded images are shown. These log-scaled images were used to check the basic symmetry of the coma and potential star traces.

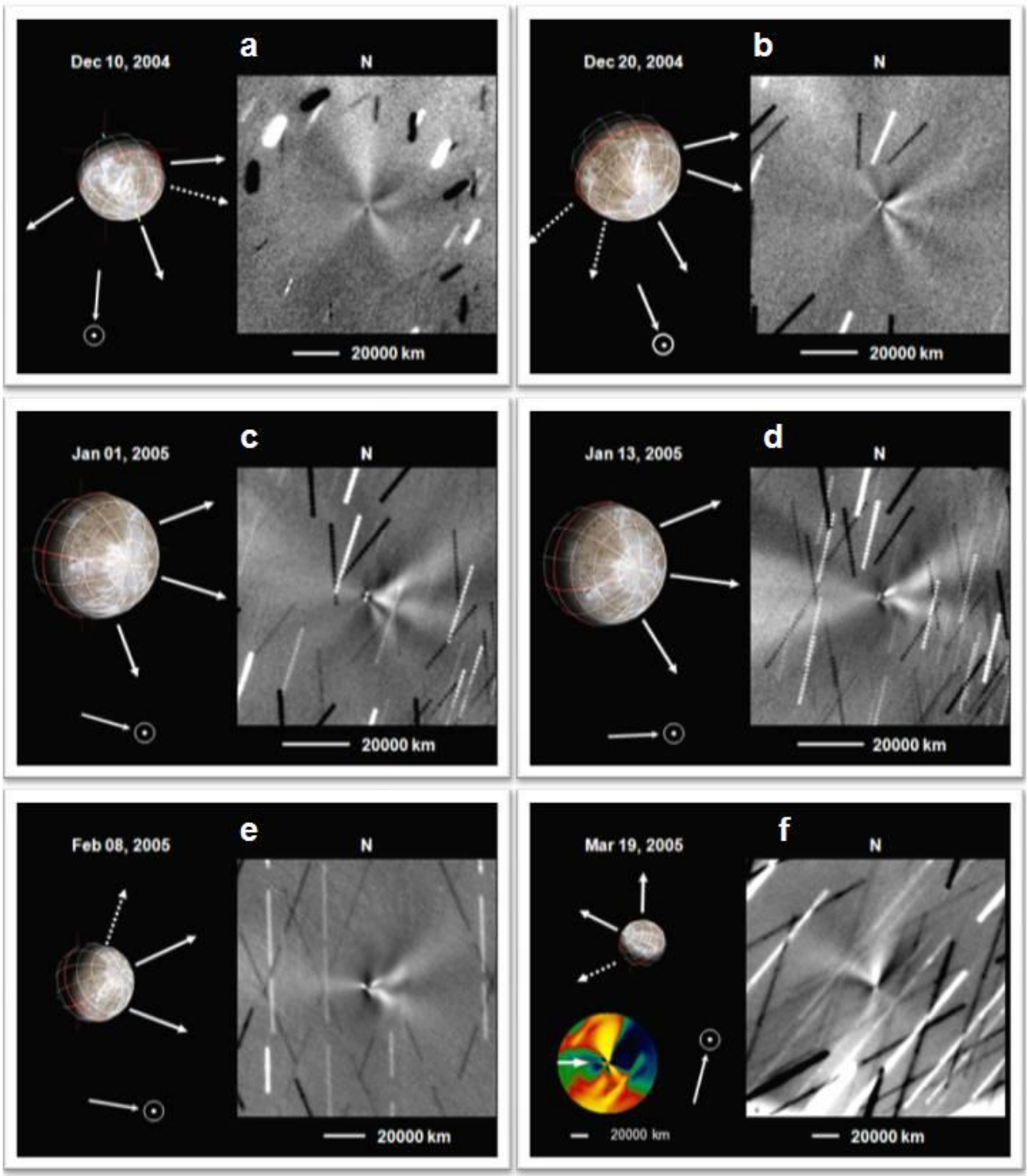

**Fig. 2 (a-f)**

The image processing of co-added images taken at different dates is shown on the right side of each panel. Rectilinear 'porcupine' jets are all around the nucleus. Black and white streaks are star trails.

A simulation of the geometrical conditions of observation is shown on the left side, with a direction of the spin axis towards RA=95° and Dec=+35°. A grid of comet-centric coordinates and the position of the rotation pole are also shown; the size of the simulations of the nucleus is scaled according to the distance from the Earth. The jet directions described in the paper are indicated with a solid arrow (jets with strong activity) or a dashed arrow (quiet jets). The direction of the Sun is also reported; the tail outflow is not shown.

**Panel f** presents also an image processed by means of the *radial de-trending* technique. A small curved jet is observable in PA 80°, hardly visible with other image processing techniques. It is likely that this structure was the symmetrical counterpart of the main jet located at latitude +45° ± 5°. Rectilinear trails are due to field stars.

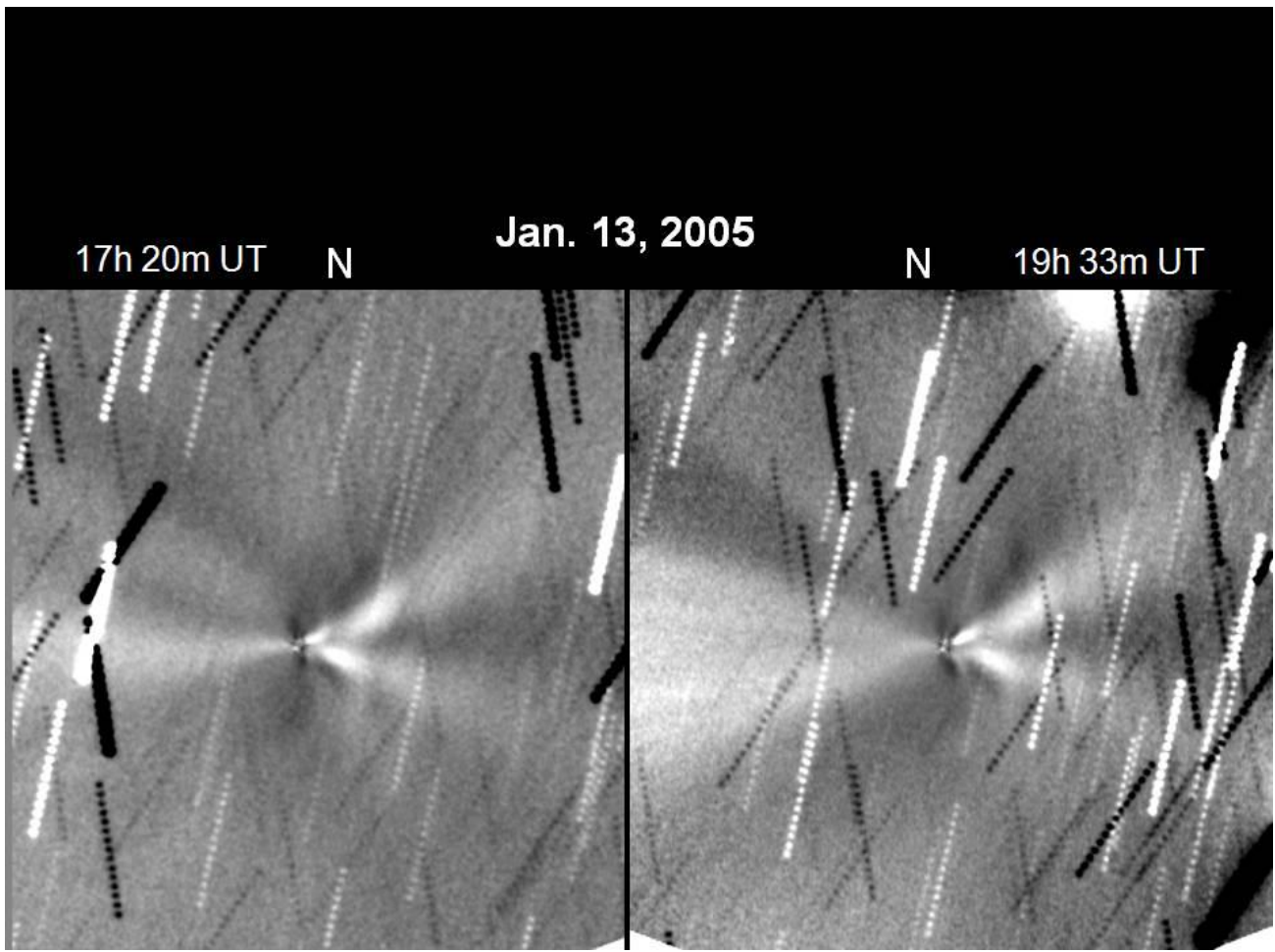

**Fig. 3**

The images taken on January 13, 2005, have been stretched to show the emission of material modulated by the rotation; the positions of "clumps" or "plumes" of material drifting away from the nucleus, fading due to expansion and gradual loss of dust, are indicated; the two images were taken only 130 minutes apart. The angle of the spin axis of the nucleus with the plane of the sky, according to our determination of the direction of the rotation axis, is about 45°.

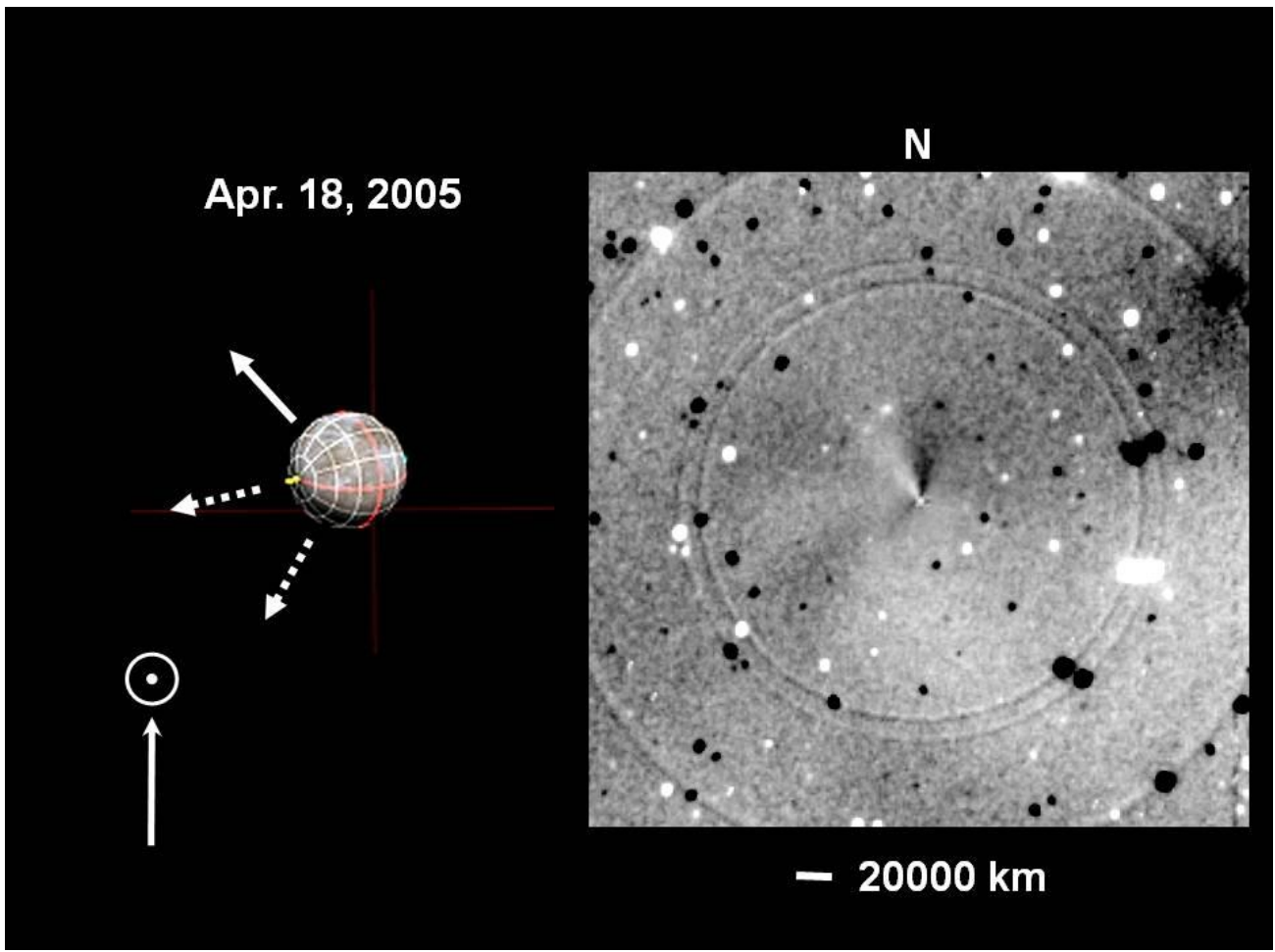

**Fig. 4**

The circular structures in the image on the right are artifacts of the processing. The geometrical simulation of the nucleus is double scaled with respect to the other figures.

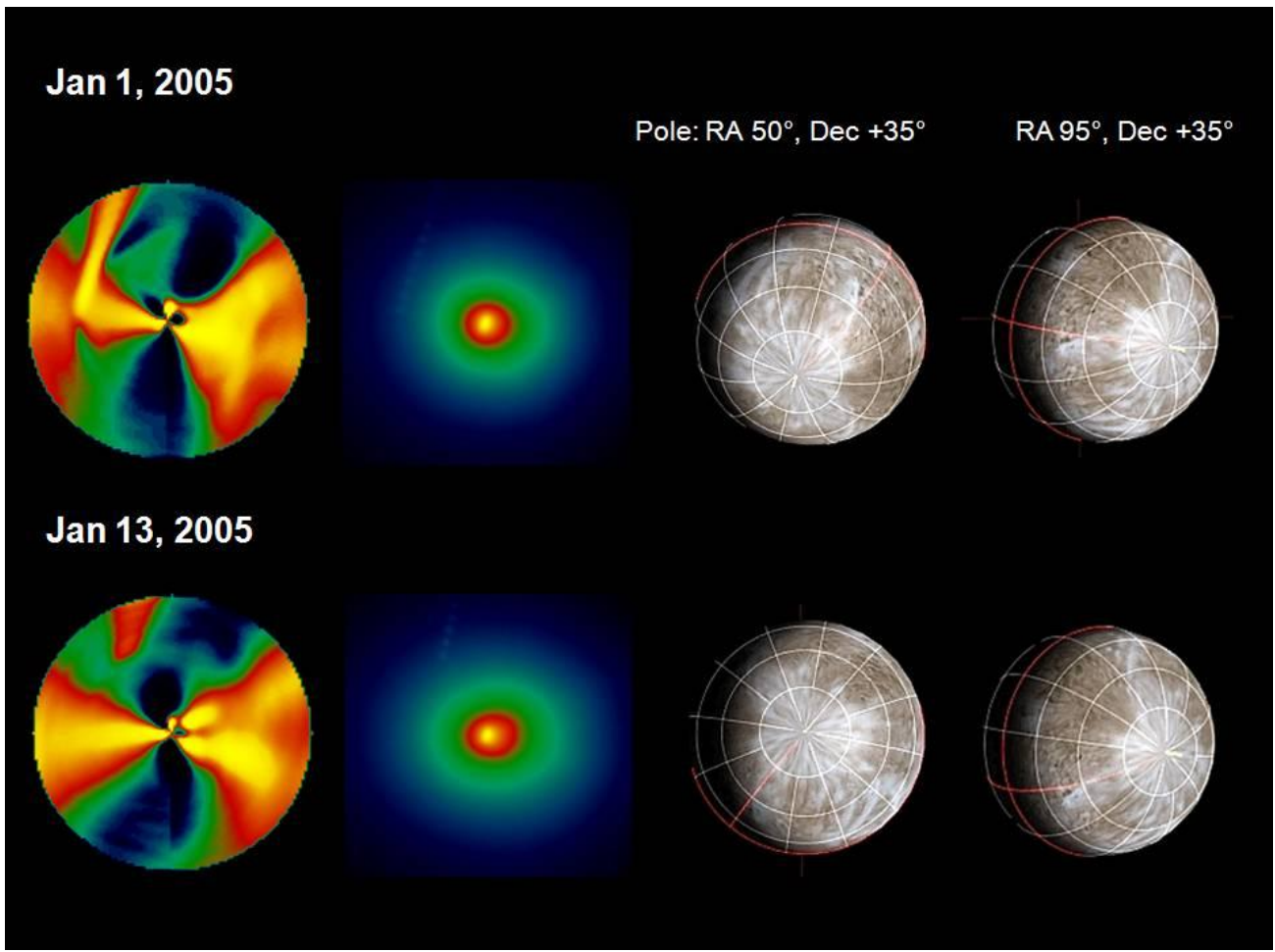

**Fig. 5**

High resolution images taken on two dates (January 1 and 13, 2005) were selected to simulate the geometrical appearance of the nucleus. The direction of the spin axis estimated from other authors (RA=50°, Dec=+35°) was also applied and, if correct, shells around the nucleus should have been observed, because the spin axis would have been oriented toward us. However, no curvilinear structures can be observed in our processed images.

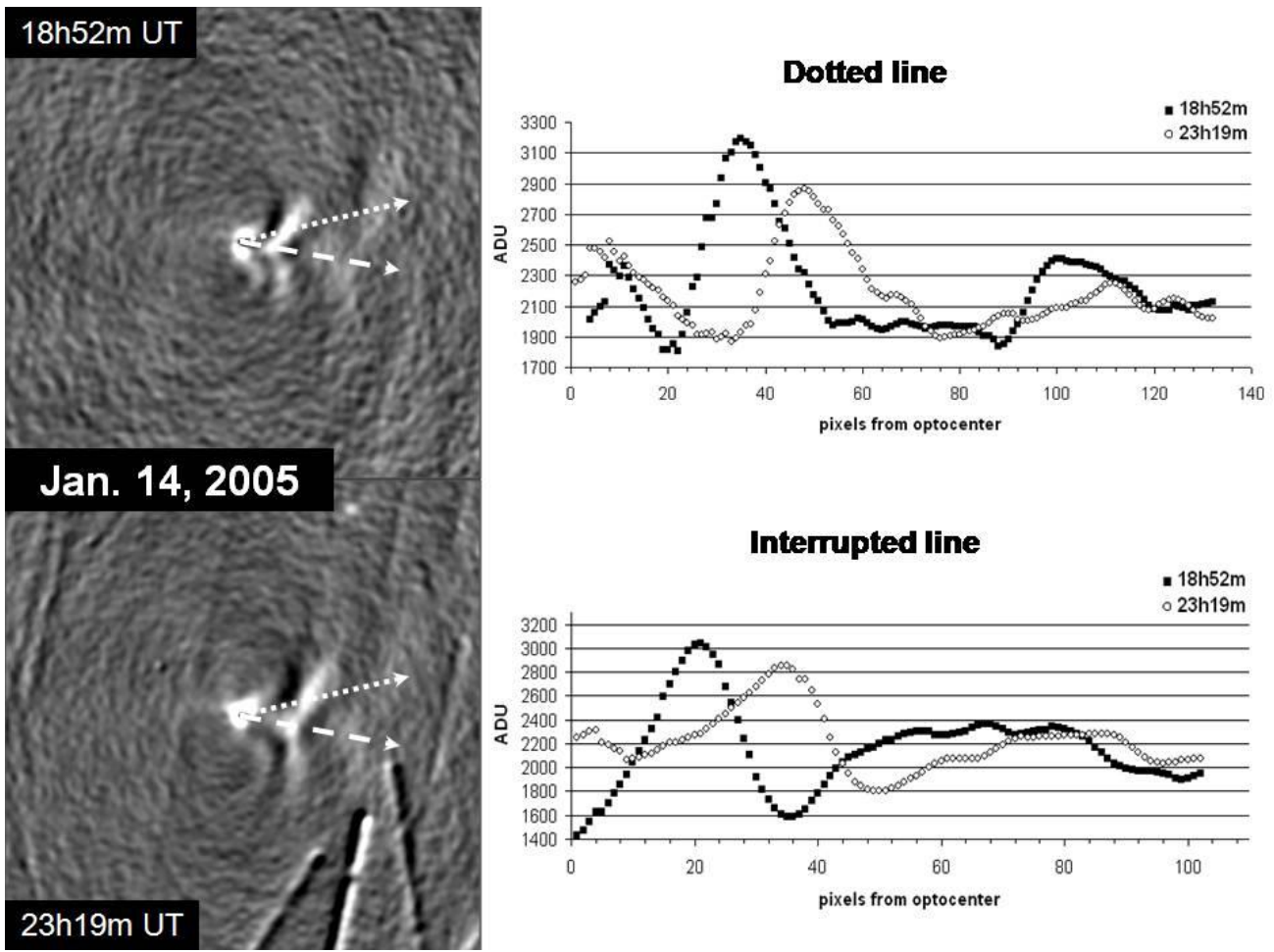

**Fig. 6**

Two images taken at on January 14, 2005 at 18h52m UT and 23h19m UT, have been magnified, processed and highly stretched and treated to better show the inner details of the coma of comet Machholz; the observed features are flowing from the nucleus of the comet. Two lines are drawn across the comet's nucleus and the brightest features of the haloes. Due to the image processing, the optocenter does not correspond with the high level of signal.

A photometric analysis of the two lines is presented on the right: the relative motion of the features allowed us to determine the rotation period of the comet's nucleus.

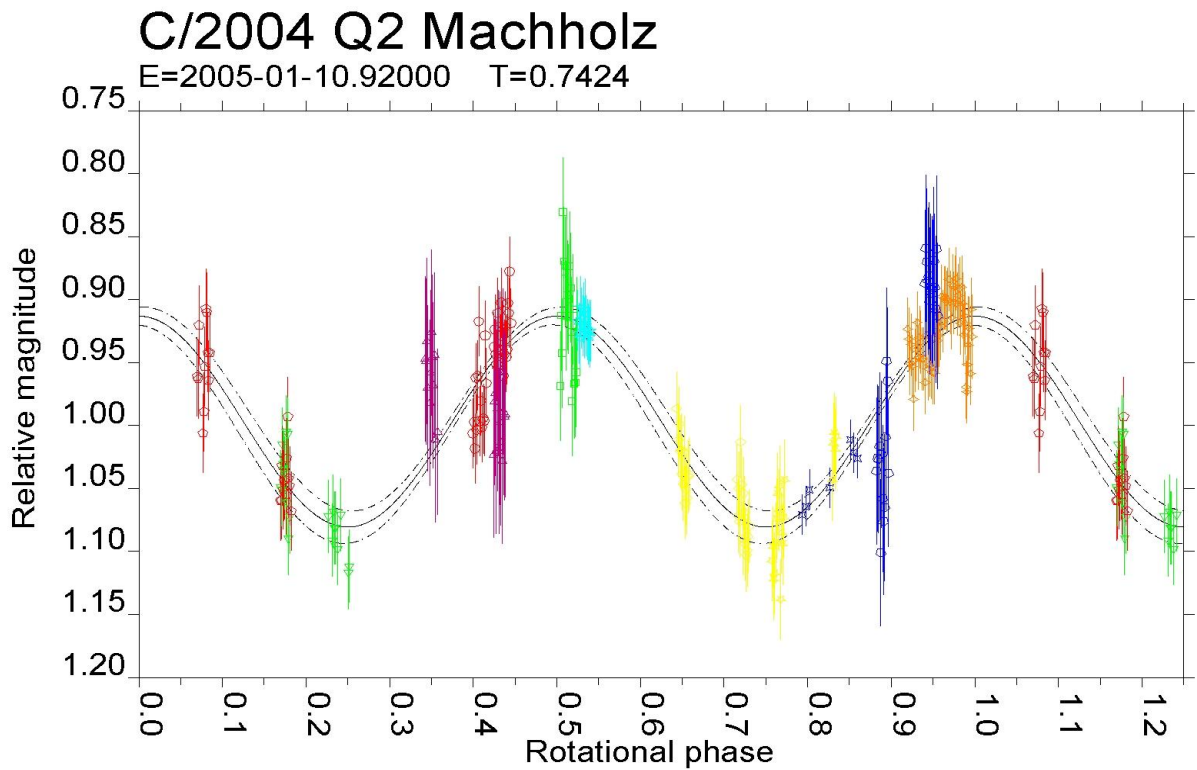

**Fig. 7**

All photometric measurements taken on images from January 3 to 14, 2005, have been merged to show the light curve of the inner central region of the coma of comet Machholz. The period (T=0.7424 d) of this curve is also the rotation period of the nucleus. Relative magnitudes are shown on the abscissa, the phases on the ordinate. Phase 0 corresponds to January 10.92000, 2005.

Each color corresponds to a single set of observations (red circles and green squares: Jan 03, 2005; blue stars and yellow diamonds: Jan 04, 2005; violet and cyan triangles: Jan 6, 2005; orange stars: Jan 10, 2005; red pentagons and green triangles: night of Jan 11, 2005; yellow stars and blue pentagons: Jan 13, 2005). Each data point is the result of the analysis of a single frame.

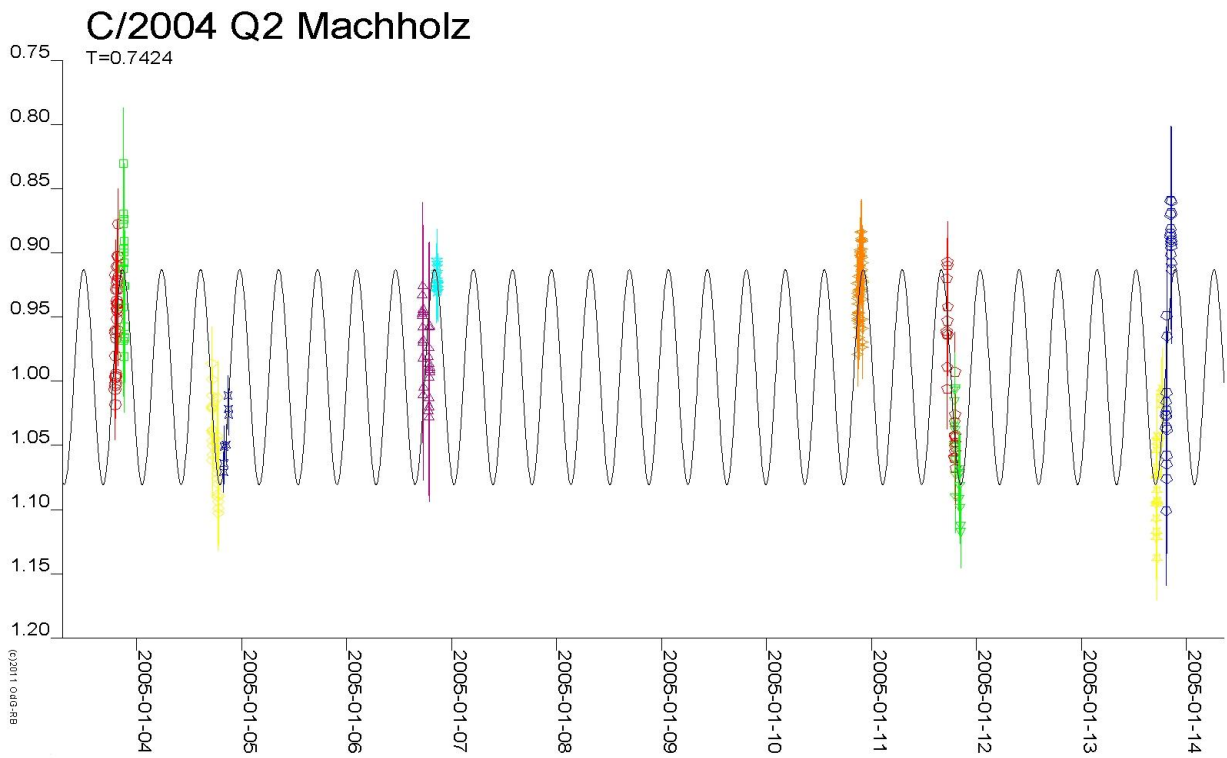

**Fig. 8**

All photometric measurements taken on images from January 6 to 14, 2005, and shown in Fig. 7 have been phased with the period (T=0.7424 d) estimated through the photometric analysis of the inner central region of the coma of comet Machholz. Relative magnitudes are shown on the abscissa, the phases on the ordinate. Phase 0 corresponds to January 10.92000, 2005. The colors indicating the different photometric measurements are the same as in Fig. 7.

**Tab. 1**

Observatories and instruments involved in the research on comet Machholz.

## Observations of Comet Machholz: Observatories

| Code | Observatory | Lat. (°) | Long. (°) | Telescope |
|------|-------------|----------|-----------|-----------|
| IAU A12 | **Stazione Astronomica Sozzago (SAS)** | | | Cassegrain 0.4m, f/6,8 |
| IAU B13 | **Tradate Observatory** | | | Simak 0.25m, f/6.4 |
| GAT#1 | **Gruppo Astronomico Tradatese** | 45.73392°N | 8.90315°E | Reflector 0.25m, f/4.8 |
| GAT#2 | **Gruppo Astronomico Tradatese** | 45.71233°N | 8.90725°E | SCT 0.2m, f/9.3 |
| IAU 106 | **Crni Crh Obs.** | | | Reflector 0.6m, f/3,4 |
| IAU A90 | **Observatorio de Sant Gervasi** | | | SCT 0.2m, f/10 |

| Code | CCD camera | Sensor | Pixel size (microns) | Filters |
|------|------------|--------|----------------------|---------|
| IAU A12 | Hi-SIS43 | Kodak KAF-1603ME | 9x9 | 700 nm broadband, center at 650 nm |
| IAU B13 | Hi-SIS43 | Kodak KAF-1603ME | 9x9 | 700 nm broadband, center at 650 nm |
| GAT#1 | SBIG ST7 | Kodak KAF-0401E | 9x9 | 700 nm broadband, center at 650 nm |
| GAT#2 | FLI ME | Kodak KAF-0401E | 9x9 | no filters |
| IAU 106 | FLI ME | TK | 24x24 | broadband |
| IAU A90 | Starlight Express MX516 | Sony | 9.8x12.6 | broadband |

**Tab. 2**

Technical data for images taken on comet Machholz in the observation sessions described in the text. The complete list of all our observations is available on request.

### Images of Comet C/2004 Q2 (Machholz)

| Date | Observ. | Mean time UT | Images during obs. sessions (N) | Exp. (s) | Atmospheric conditions | FWHM of stars (pix) | Scale ("/pix) | Scale (km/pix) | $\Delta$ (AU) | $\Delta$ (10^6 km) | R (AU) | R (10^6 km) | Elong. (°) | Phase angle (°) |
|---|---|---|---|---|---|---|---|---|---|---|---|---|---|---|
| 2004-12-10 | IAU A12 | 22.13 | 15 | 60 | good | 2,8 | 1,36 | 501 | 0,508 | 76,0 | 1,398 | 209,1 | 135 | 30,0 |
| 2004-12-10 | GAT#1 | 22.28 | 100 | 30 | good | 2,4 | 1,51 | 556 | 0,508 | 76,0 | 1,391 | 208,2 | 135 | 30,0 |
| 2004-12-30 | GAT#2 | 21.36 | 30 | 60 | good | 2,4 | 1,41 | 365 | 0,357 | 53,4 | 1,267 | 189,5 | 137 | 32,3 |
| 2005-01-01 | IAU A12 | 21.30 | 20+25 | 60 | very good | 2,1 | 1,36 | 346 | 0,351 | 52,5 | 1,257 | 188,0 | 135 | 33,5 |
| 2005-01-02 | IAU A12 | 20.35 | 15+20 | 60 | fair | 2,8 | 1,36 | 344 | 0,349 | 52,2 | 1,253 | 187,4 | 134 | 34,2 |
| 2005-01-03 | GAT#2 | 22.20 | 50 | 60 | very good | 2,9 | 1,41 | 356 | 0,348 | 52,1 | 1,249 | 186,8 | 133 | 34,8 |
| 2005-01-04 | GAT#2 | 21.26 | 50 | 60 | good | 3 | 1,41 | 355 | 0,347 | 52,0 | 1,245 | 186,3 | 132 | 35,6 |
| 2005-01-06 | IAU A12 | 19.15 | 20+20+20 | 60 | very good | 2,5 | 1,36 | 343 | 0,348 | 52,1 | 1,237 | 185,1 | 130 | 36,4 |
| 2005-01-10 | IAU A90 | 21.15 | 57 | 60 | good | 2,4 | 1,10 | 320 | 0,355 | 53,5 | 1,224 | 183,1 | 126 | 40,5 |
| 2005-01-11 | IAU A12 | 19.20 | 15+16+15 | 60 | good | 1,9 | 1,36 | 353 | 0,358 | 53,6 | 1,222 | 182,8 | 124 | 41,6 |
| 2005-01-13 | IAU A12 | 19.40 | 15+15+15 | 60 | very good | 1,9 | 1,36 | 360 | 0,365 | 54,6 | 1,217 | 182,1 | 122 | 43,3 |
| 2005-01-13 | IAU B13 | 19.30 | 60 | 60 | good | 2,6 | 1,14 | 302 | 0,365 | 54,6 | 1,217 | 182,1 | 122 | 43,3 |
| 2005-01-14 | GAT#1 | 21.10 | 50+95 | 30 | good | 2,2 | 1,51 | 405 | 0,370 | 55,3 | 1,215 | 181,8 | 121 | 44 |
| 2005-02-08 | IAU B13 | 23.50 | 360 | 60 | good to fair | 2,8 | 1,14 | 474 | 0,573 | 85,7 | 1,228 | 183,7 | 100 | 52,2 |
| 2005-03-19 | IAU A12 | 21.15 | 200 | 60 | good to fair | 2,3 | 1,36 | 953 | 0,966 | 144,5 | 1,464 | 219,0 | 96 | 42,5 |
| 2005-04-18 | IAU 106 | 01.12 | 4 | 90 | good | 1,6 | 2,49 | 2326 | 1,288 | 192,7 | 1,749 | 261,6 | 98,6 | 34,6 |

**Tab. 3**

Geometric conditions of the appearance of comet Machholz during the photometric sessions used for the determination of the rotation period.

### Images of Comet C/2004 Q2 (Machholz)
Geometric conditions during photometric sessions

| Date UT | D (UA) | R (UA) | l (°) | b (°) | l_PAB (°) | b_PAB (°) | Phase (°) | Symbols on figures |
|---------|--------|--------|-------|-------|-----------|-----------|-----------|--------------------|
| 2005-01-03.8 | 0.3483 | 12.492 | 92.0 | 1.4 | 74.7 | -3.3 | 34.7 | red circles and green squares |
| 2005-01-04.8 | 0.3475 | 12.455 | 92.7 | 0.8 | 75.0 | -1.9 | 35.5 | blue stars and yellow diamonds |
| 2005-01-06.8 | 0.3474 | 12.379 | 94.3 | 0.5 | 75.7 | 1.1 | 37.1 | violet and cyan triangles |
| 2005-01-10.9 | 0.3544 | 12.249 | 97.5 | 3.0 | 77.6 | 7.3 | 40.7 | orange stars |
| 2005-01-11.8 | 0.3571 | 12.224 | 98.2 | 3.6 | 78.0 | 8.6 | 41.4 | red pentagons and green triangles |
| 2005-01-13.8 | 0.3646 | 12.176 | 99.9 | 4.9 | 79.1 | 11.5 | 43.1 | yellow stars and blue pentagons |

**D** = Comet-Earth distance
**R** = Sun-Comet distance
**l** and **b** = heliocentric ecliptical longitude and latitude of the comet
**l_PAB** and **b_PAB** = heliocentric ecliptical longitude and latitude of the phase angle bissector
**Phase** = Sun-Comet-Earth angle